# Electric Switching of Fluorescence Decay in Gold-Silica-Dye Nematic Nanocolloids Mediated by Surface Plasmons

Li Jiang,[†] Haridas Mundoor,[†] Qingkun Liu,[†] and Ivan I. Smalyukh[*,†,‡,§]

[†]*Department of Physics and Soft Materials Research Center, University of Colorado, Boulder, CO 80309, USA*

[‡]*Department of Electrical, Computer, and Energy Engineering, Materials Science and Engineering Program, University of Colorado, Boulder, CO 80309, USA*

[§]*Renewable and Sustainable Energy Institute, National Renewable Energy Laboratory and University of Colorado, Boulder, CO 80309, USA*

*\* Correspondence to: ivan.smalyukh@colorado.edu*

## ABSTRACT

Tunable composite materials with interesting physical behavior can be designed through integrating unique optical properties of solid nanostructures with the facile responses of soft matter to weak external stimuli, but this approach remains challenged by their poorly controlled co-assembly at the mesoscale. Using scalable wet chemical synthesis procedures, we fabricated anisotropic gold-silica-dye colloidal nanostructures and then organized them into the device-scale (demonstrated for square inch cells) electrically tunable composites by simultaneously invoking molecular and colloidal self-assembly. We show that the ensuing ordered colloidal dispersions of shape-anisotropic nanostructures exhibit tunable fluorescence decay rates and intensity. We characterize how these properties depend on low-voltage fields and polarization of both the excitation and emission light, demonstrating a great potential for the practical realization of an interesting breed of nanostructured composite materials.

Surface plasmon resonance (SPR) effects continue to attract a great deal of applied and fundamental interest and impinge on diverse research fields including nanotechnology, biophysics, nanophotonics, materials science, biotechnology, nanofabrication and many other areas.[1-4] These effects are capable of re-defining the nature of light-matter interactions while also localizing them within spatially confined regions much smaller than wavelength of light. These SPR effects have been exploited in enhancing weak fluorescence and Raman signals for biological imaging and detection,[5,6] in altering photo-conversion processes exhibited by quantum dots and upconversion particles,[7,8] in enabling photothermal conversion,[9] in realizing nanolasers and related physical phenomena,[10-13] as well as for a variety of other research and technological needs. The SPR effects are important for the realization of a broad variety of optical metamaterials with pre-designed optical properties, such as the effective medium refractive index.[14] Perhaps the most interesting but yet largely unexploited possibilities arise in the design and realization of tunable emissive materials in which the SPR-related losses of light can be compensated by the corresponding enhancement effects. For example, a colloidal dispersion of metal-silica-dye nanostructures can be pre-engineered to demonstrate the surface plasmon amplification by stimulated emission of radiation (SPASER) with a host of potential practical uses.[10] There is a great demand in developing means of controlling the SPR-based light-matter interactions that lead to such profound effects using weak external stimuli, such as low-voltage fields and light itself. On the other hand, realization of reconfigurable composite materials with tunable properties arising from such SPR-enabled light-matter interactions has the potential of revolutionizing a broad range of technologies.

We develop, realize and characterize a composite material with reconfigurable fluorescence properties, in which both the intensity and the decay rates of fluorescence can be controlled by electric fields ~1 V. These composite materials combine pre-engineered rod-like gold-silica-dye (GSD) colloidal nanostructures and a liquid crystal (LC) host medium that imposes orientational ordering on these nanorods and that mediates synchronous switching of anisotropic particle orientations within it. The fluorescence of dye molecules randomly distributed within silica shells of these core-shell GSD nanorods is then altered through the polarization- and voltage-dependent Purcell effect to mediate facile control of the composite's emissive properties. To assure the broad applicability and technological potential of our approach, we have used a well-known commercially available liquid crystal host. This approach of realizing highly tunable and reconfigurable properties of composite materials by combining SPR-enabled nanoscale interactions with the facile response and intrinsic orientational ordering of the LC host can be potentially extended to enable tunable nanolasers and quantum-dot-based single-photon-emitters.

**RESULTS AND DISCUSSION**

**Composition and Properties of GSD Nanostructures.** The hybrid GSD nanostructure is shown schematically in Figure 1a. The Transmission Electron Microscopy (TEM) image of GSD nanocomposites in Figure 1b shows uniform silica capping around gold nanorods (GNRs). The average length and diameter of GNRs are 47.8 nm and 20.4 nm, respectively, while the average thickness of silica layer is 10.2 nm, as determined using multiple TEM images (based on size distributions of GNRs, Figure S1). To get insights into the photo physical properties of such GSDs, we simulated the electric field distribution around the gold nanoparticles within them, which was calculated at wavelengths corresponding to transverse (530 nm) and longitudinal (660 nm) SPR modes of the GNR and depicted in Figure 1c-f. Gold nanoparticles enhance the electric field around them because of their characteristic SPR modes. The anisotropic shape of the GSD nanostructures makes these properties of individual particles polarization dependent. The particle was designed to produce strong field enhancement when the polarization of the incident light (**P**) is either parallel to the short axis of the GSDs at excitation wavelength (Figure 1c) or parallel to the long axis of the rods at emission wavelength (Figure 1f), which is expected to affect the fluorescence emission of the silica-embedded dye molecules in the following way. The field enhancement at the excitation wavelength is expected to induce an increase in the absorption rates of the dye molecules in the silica-shell part of the GSD. On the other hand, the emission from these dye molecules was pre-designed to couple strongly with the gold nanoparticles to provide an increase in the fluorescence decay rates through the Purcell effect. The strengths of such optical interactions are expected to depend on the distance of individual dye molecules from gold core and their relative orientation with respect to the rod-like gold nanoparticles, owing to the absorption and fluorescence anisotropy of the individual dye molecules as well as the properties of SPR effects. In order to explore the possibility of utilizing such interactions in engineering properties of nanostructures, we characterized them through optical experimentation, as described below.

**Alignment and Electric Switching of GSDs in a Nematic LC.** After coating a silica layer, the longitudinal SPR peak of GSDs red-shifts by ~14 nm (Figure 2c, top) as compared to the bare GNRs because of the mesoporous silica shell's higher refractive index (1.45) as compared to that of water (1.33). GSDs treated with

Dimethyloctadecyl[3-(trimethoxysilyl)propyl]ammonium chloride (DMOAP) produce perpendicular surface boundary conditions (anchoring) for the LC director **n** describing the local average orientation of the constituent rod-like organic LC molecules (Figure 2a). The physical behavior of such composites is largely dependent on the competition between the free energy costs associated with the surface free energy due to the deviation of the director from the "easy-axis" orientations prescribed by the surface anchoring conditions at the nanoparticle-LC interface and the bulk elastic free energy costs due to distortions of the uniform director alignment caused by these boundary conditions on the particle surface. This competition is commonly quantified in terms of the so-called "surface anchoring extrapolation length" $l_e = K/W$, where $K$ is the average Frank elastic constant of the LC host and $W$ is the surface anchoring coefficient characterizing the energetic costs for the deviation of **n** from the easy axis at the LC-nanoparticle interface. When nanoparticles of size comparable or smaller than $l_e =100$-$500$ nm are incorporated into the LC host, like in the case of our GSDs, the weak-anchoring-like behavior of the LC-based nanocomposite is expected.[15] In particular, since the distortions of the director field around the particle are energetically costly, one expects significant deviations of the director from the easy-axis orientations at the particle surfaces, albeit weak bulk distortions of the director can be present too. To minimize the energetic costs associated with surface anchoring and bulk elastic free energies due to the rod-like nano-inclusions in the aligned LC host, the GSDs spontaneously orient, on average, at an angle $\theta \approx 90°$ with respect to the far-field director **n**$_i$ while freely rotating around the uniform far-field **n**$_i$ (Figure 2b). This alignment of GSDs results in strong polarization-dependent extinction spectra, which was tested in a planar cell with the initial unidirectional LC molecular alignment along **n**$_i$ with the polarization **P** of the normally incident light rotatable in the plane of the LC cell and $n_i$. For light with $P \| n_i$ propagating orthogonally to the LC cell substrates, mostly only the transverse SPR mode of GNRs is excited, so that a relatively small longitudinal SPR extinction peak is observed at around 680nm (Figure 2c). Both transverse and longitudinal SPR peak of GNRs red-shift with increasing the refractive index of medium and the longitudinal SPR peak is more sensitive to the refractive index of the host medium than the transverse peak for GNRs, consistent with our previous experimental and numerical modeling studies.[16] For $P \perp n_i$, a mixed excitation of transverse and longitudinal SPR modes is observed, with a rather strong extinction at longitudinal SPR peak (Figure 2c, bottom), consistent with the fact that the average nanorod orientation is perpendicular to $n_i$.

To quantify the LC-host-induced orientational ordering of GSDs, we determine the scalar order parameter characterizing the ordering of nanoparticles and defined as $S_{GSD} = (3\langle \cos^2 \theta \rangle - 1)/2$, where $\theta$ is the angle between the long axes of GSDs and $n_i$ and the angle brackets $\langle \ \rangle$ denote averaging over orientations of the GSDs. In general, when defined in this way, $S_{GSD}$ can vary from -1/2 to +1. For a sample, for completely disordered orientations of GSDs, $S_{GSD}= 0$, whereas $S_{GSD}= 1$ for a perfectly ordered sample with all GSDs in the dispersion oriented along $n_i$ and $S_{GND}= -1/2$ for a sample with all GSDs oriented perfectly in directions perpendicular to $n_i$. Because of the perpendicular surface boundary conditions for the director on nanoparticle surfaces, our GSDs tend orient perpendicular to $n_i$, though the thermal fluctuations tend to deviate them away from these free-energy-minimizing orientations, so that the value of $S_{GND}$ is expected to be negative but with absolute value smaller than 1/2. Indeed, using our extinction spectra measured for polarizations $P\|n_i$ and $P\perp n_i$, the scalar orientational order parameter of GSDs is determined as $S_{GSD} = (Extinct_{\|} - Extinct_{\perp})/(Extinct_{\|} + 2Extinct_{\perp}) = -0.28$, where $Extinct_\parallel$ and $Extinct_\perp$ are the extinction values at the longitudinal SPR peak for $\mathbf{P}\|\mathbf{n_i}$ and $\mathbf{P}\perp\mathbf{n_i}$, respectively.

The LC-mediated alignment of GSDs can further be switched electrically. We have used a commercially available nematic LC 4-cyano-4'-pentylbiphenyl (5CB), which has a positive dielectric anisotropy $\Delta\varepsilon=\varepsilon_\parallel-\varepsilon_\perp=9$, where $\varepsilon_\parallel$ and $\varepsilon_\perp$ are dielectric constants measured for fields parallel and perpendicular to $\mathbf{n}$, respectively. Thus, voltages applied to a planar cell produce a vertical electric field, so that $\mathbf{n}$ rotates to become perpendicular to the cell plane and both transverse and longitudinal SPR modes are excited at all linear polarizations of incident light. Figure 2c (bottom) shows that the extinction coefficient is relatively low at the wavelength of the longitudinal SPR peak at $\mathbf{P}\|\mathbf{n_i}$ and at no applied fields, but it can be "switched on" by applying voltage to the indium tin oxide (ITO) electrodes. In contrast, the extinction spectra barely change (Figure 2c, bottom) when the electric field is applied for $\mathbf{P}\perp\mathbf{n_i}$. Thus, the SPR modes can be effectively controlled by changing polarization of light and by electric fields, which we will utilize for tuning the fluorescence properties of the GSD nanostructures.

The long-range alignment of GSDs with negative scalar order parameter emerges mainly from minimizing surface free energy with respect to the relative orientations of the uniform $\mathbf{n}$ and LC-GSD interfaces with finite homeotropic boundary conditions. Because our GSDs are smaller than the surface anchoring extrapolation length $l_e = 100\text{-}500$ nm, the director couples only weakly to GSD surfaces, so that elastic colloidal interactions can be neglected (in the case of strong distortions of $\mathbf{n}$ around particles, the GSDs would interact to reduce the overall distortions of the uniform alignment of the LC to minimize free energy[16]). The surface anchoring energy is minimized when short axes of nanorods align along $\mathbf{n}$ and the long axes orthogonally to it, although thermal fluctuations tend to randomize their orientations. GSD is modeled as an ellipsoid-shaped particle with short axis $a = 40.8$ nm and long axis $b = 68.2$ nm (based on size distributions of GSDs, Figure S1). Assuming the Rapini−Papoular form of the surface anchoring potential, $f_{sa} = (W/2)\sin^2\psi$, where $\psi$ is an angle between **n** and the "easy" axis normal to the GSD's surface,[17] we express $\psi$ in terms of polar angle $\beta$ and azimuthal angle $\phi$ and integrate the free energy density $f_{sa}$ over the entire surface area of the ellipsoid in the spherical coordinate system. We then find the total surface free energy due to the GSD through numerical integration:

$$F_{sa}(\theta) = \frac{1}{2}W\int_0^{2\pi}\int_0^{\pi}\sin\beta\left(a^2\sin^2\beta + b^2\cos^2\beta\right)\left(1 - \frac{(b\sin\beta\sin\phi\sin\theta + a\cos\beta\cos\theta)^2}{b^2\sin^2\beta + a^2\cos^2\beta}\right)d\beta d\phi$$

where $W$ is the polar surface anchoring coefficient and $\theta$ is the angle between $\mathbf{n_i}$ and the long axis of the ellipsoid. The equilibrium distribution of ellipsoid orientations due to these interactions follows Boltzmann statistics $f_s(\theta) \propto \exp[-F_{sa}(\theta)/k_B T]$, where $k_B$ is Boltzmann constant, and $T \approx 300$ K is the absolute temperature. The orientational order parameter can be found as[18] $S_{GSD} = \int_0^{\pi} P_2(\cos\theta)f_s(\theta)\sin\theta d\theta$, where $P_2$ denotes the second Legendre polynomial. For geometric parameters of GSDs and experimentally measured scalar order parameter corresponding to the angular distribution of GSD orientations shown in Figure 2d, we estimated $W = 4.6 \times 10^{-5}$ J/m$^2$, which is in a good agreement with the values independently determined in the previous studies.[15] The orientational ordering of GSDs with negative scalar order parameter depends on the geometry of the GSD modeled as an ellipsoid and the surface anchoring strength. The inset of Figure 2d shows how $S_{GSD}$ can be tuned from $S_{GSD} = 0$ to $S_{GSD} \approx -0.5$ by changing $W$ for the experimental aspect ratio of the used GSDs or by changing aspect ratio ($b/a$) at the fixed $W = 4.6 \times 10^{-5}$ J/m$^2$. This demonstrates the means of controlling the spontaneous orientational order of GSDs, as needed to tune their fluorescence properties that we discuss below.

**Surface-Plasmon-Mediated Electric Switching of Fluorescence Intensity.** When the GSD-LC composite infiltrated into planar cells is excited by green light (510-560 nm) from the supercontinuum source, the GSD particles fluoresce, producing diffraction limited bright spots undergoing Brownian motion, as displayed in Figure 3a. Figure 3b shows the fluorescence spectrum of GSD nanocomposites, with the inset depicting the normalized fluorescence intensities detected under different combinations of linear polarization of incident excitation light **P**, the analyzer **A** in the fluorescence detection channel, and applied voltage $U$. Within the excitation wavelength range of 510-560 nm, the absorption of the dye would be inefficient without the field enhancement induced by the transverse SPR mode. The fact that the transverse SPR enhancement is so strong that it makes it efficient is one of the advantages of our approach, allowing us to substantially separate the excitation and emission wavelengths and determine the underpinnings of the studied physical behavior. At **P**||$n_i$ (orthogonal to ellipsoidal GSDs), the gold nanoparticles induce strong excitation field enhancement through the transverse SPR mode. The experimental geometry with an analyzer kept parallel to $n_i$ allows us to detect the enhanced fluorescence signal from the dye molecules, which are roughly co-aligned with gold nanoparticles with maximum absorption of the enhanced excitation from GSDs, resulting in an enhanced fluorescence emission. This particular configuration with **P** and **A** along $n_i$ and along the short axes of all GSDs produces maximum fluorescence intensity from the nanocomposites, as envisaged by our design. When the analyzer is rotated by 90°, fluorescence detected through the analyzer is mostly from the dye molecules oriented perpendicular to the incident polarization, which has the minimum fluorescence enhancement due to their orientation with respect to the gold nanoparticle, resulting in a lower fluorescence intensity. A Cosine-square-like variation is observed in the fluorescence intensity when **A** is rotated by 180° with respect to $n_i$ while keeping

**P** along **n_i**, as shown in Figure 3c. When voltage $U = 10$ V is applied to switch the LC director from in-plane to vertical orientation, the ellipsoidal particles rotate synchronously with **n**, yielding an isotropic distribution of in-plane GSD orientations orthogonal to the direction of the incidence of light. Consequently, the overall fluorescence intensity becomes lower (Figure 3b) and independent of the orientations of **P** and **A**. This polarization-dependent switching of fluorescence enhancement is consistent with all used **P**- and **A**-orientations with respect to the rubbing direction for $U = 0$ V and 10 V (Figure 3b).

The fluorescence of GSDs in the LC medium can be controlled continuously at voltages as low as 1 V. To demonstrate this, we measured the variation of fluorescence intensity with voltage $U$ at both **P** and **A** parallel to **n_i** for the composite of GSDs in 5CB in a planar cell (Figure 3d). The fluorescence intensity of light emitted by GSDs is maximum at 0 V, decreases when the applied voltage exceeds ~1 V threshold for the LC director switching and eventually saturates at higher voltages, revealing a threshold-like variation in the intensity, with the threshold voltage $U_{th} \approx 1$ V indicated by the dashed line in Figure 3d. Similar to pristine LCs and other LC-nanoparticle composites we studied previously, the LC-GSD composites do not respond to applied voltages below the threshold value, but readily switch at voltages above the threshold determined by the competition of elastic and electric coupling terms of the free energy,[16] which is commonly referred to as "electric-field-induced Fredericks transition". Although GSDs at used concentrations can slightly modify LC properties, the ~1 V realignment threshold of our composite is close to $U_{th} = \pi[K_{11}/(\varepsilon_0 \Delta\varepsilon)]^{1/2}$ of the pristine 5CB host,[18] where $\varepsilon_0 = 8.854 \times 10^{-12}$ CV$^{-1}$m$^{-1}$ and $K_{11} = 5.4$ pN is the splay elastic constant of 5CB. The nature of switching of the GSD-LC composites was further analyzed by measuring the fluorescence intensity while switching the voltage on and off and increasing $U$ in steps of 1 V (Figure 3e). The fluorescence intensity (Figure 3e) is modulated in accordance with the applied voltage. The director realignment dynamics can be characterized by ON and OFF times that are deduced from transmission *versus* time curves at different switching voltages for **P**∥**n$_i$** by measuring the time span for transmission intensity changes between 10% and 90% of the original value (Figure 3f).[16] For pristine 5CB LC, the theoretical value of OFF time can be determined as $\tau_{OFF} = \gamma_1 d^2 / (K_{33} \pi^2)$,[18] where $\gamma_1$ is the rotational viscosity, $d$ is thickness of the LC cell, and $K_{33}$ is the bend elastic constant. For pristine 5CB, $\gamma_1$=0.0806 Pas and $K_{33}$=7.2 pN, resulting an OFF time of 1.02 s for a 30 µm cell. For the GSD-5CB composite, the measured OFF time is around 2.78 s. The possible explanations for this slower response include the increased rotational viscosity of the composite as compared to pristine 5CB and time lagging between reorientation of particles as compared to the rotation of director of the LC host.[15] The ON time decreases with U, qualitatively similar to the case of pristine and other types of guest-host LCs.[15,18]

**Surface-Plasmon-Mediated Electric Switching of Fluorescence Lifetime.** In Figure 4a, we compare the fluorescence decay characteristics of the dye Oxazine 725 doped directly into the LC matrix and that of the GSD-LC composite, in which the same dye molecules are entrapped in the silica matrix surrounding the GNRs. The fluorescence decay data for the dye doped LC sample show single-exponential decay curves with the characteristic decay time 2.58-2.87 ns, comparable to the lifetime of Oxazine dye molecule in the silica composite ~2.51 ns,[19] being relatively independent of the orientation of **A** with respect to **n$_i$**. For the GSD-LC composites, however, the fluorescence follows a double exponential decay and this fluorescence decay is the fastest at **A**⊥**n$_i$** (Figure 4a). Figure 4b represents the fluorescence decay plots for the GSDs-LC composite when it is switched by $U$ = 10 V, for **P** and **A** oriented either parallel to the initial orientation of the far-field **n$_i$** at $U$ = 0 or perpendicular to it. Variation of fluorescence decay with switching of the composite is evident, which for the **A**∥**n**$_i$ case is independent of the polarization of the excitation beam. Further to this, we have characterized the fluorescence decay curves while increasing $U$ from 0 to 10 V with the increment of 1 V for **P**, **A**∥**n**$_i$ (Figure 4c), which is found to be consistent with the same pre-designed polarization-dependent fluorescence properties of the GSD-LC composite.

To further demonstrate electric control of the fluorescence decay, we have measured the time that corresponds to the decay to 10% of the initial fluorescence counts, as shown in Figure 4d. We observed a threshold-like dependence of this time on voltage, with the threshold voltage value ~1 V comparable to that derived from probing fluorescence intensity (Figure 3d). By fitting the experimental data with a double exponential curve, $I = I_0 + A_1 e^{[-(t-t_0)/\tau_1]} + A_2 e^{[-(t-t_0)/\tau_2]}$, we extracted two characteristic life times describing a faster decay component with $\tau_1 \approx$ 0.47-0.61 ns and slower decay component with $\tau_2 \approx$ 2.16-2.87 ns. Although the ratio $\tau_1/\tau_2$ remains essentially independent of $U$ (inset of Figure 4d), the relative values of amplitudes $A_1$ and $A_2$ of the fitting exponents systematically vary with $U$ and also exhibit threshold-like behavior (Figure 4e).

The fluorescence properties of dye molecules in the proximity of plasmonic nanoparticles and within plasmonic nanostructures, similar to our GSD particles, were studied previously.[19-22] It was shown that these fluorescence properties are strongly influenced by the SPR effects.[19-22] However, to the best of our knowledge, such SPR-fluorescence interplay has not been exploited as a means of controlling physical behavior of nanostructured composites. In our specially pre-engineered colloidal GSD nanostructures, the emission wavelength of the dye molecules is matched to the longitudinal SPR mode of the gold nanoparticles in the GSD core. This pre-engineering aims at the SPR-enabled controllable modification of fluorescence decay rates of the dye molecules as the anisotropic GSD particle orientations are switched. The relative change in the radiative decay rate of a dye molecule can be characterized by the so-called Purcell factor $F_p = 8/(3\pi^2)(\lambda/n)^3 Q/v$, where $Q$ is the quality factor, $v$ is the mode volume, $n$ is refractive index of the medium and $\lambda$ is the emission wavelength.[21] Strong enhancement of the electromagnetic field due to the SPR results in lower values of $v$ and, hence, in an increase of the overall value $F_p$ (by a factor of ~150), as reported earlier.[21] The strength of such interaction is dependent on both positions and orientations of these dye molecules within the silica shell, albeit they are distributed at random, both in terms of positions and orientations, and do not have any preferential orientation with respect to the GNR core. The rod-like dye molecules with their emission polarized along the long axis of the GNRs are subjected to the maximum change in the fluorescence decay rate, which corresponds to the faster decay component with a characteristic time $\tau_1 \approx 0.47$-$0.61$ ns (Figure 4). On the other hand, the fluorescence decay rates for the dye molecules with emission polarizations parallel to short axis of the gold nanoparticles are relatively unaffected by the Purcell effect, retaining $\tau_2 \approx 2.16$-$2.87$ ns close to that of the dye molecules dispersed directly in the LC or in silica microspheres without GNRs.[19] Thus, the experimentally measured fluorescence decay curves presented in Figure 4 can be interpreted as the superposition of the faster and slower decay curves, with the coefficients $A_1$ and $A_2$ dependent on the relative orientation of the dye molecules with respect to the rod-like gold cores and their positions with respect to the GNR surface. The fact that the distribution of dye molecule orientations stays random at all studied conditions allows us to demonstrate that the effective control of fluorescence decay characteristics emerges purely from the polarization- and voltage-dependent Purcell effect.

To demonstrate a composite material with tunable florescence properties we exploit the fact that GSDs spontaneously align nearly perpendicular to $\mathbf{n_i}$ when dispersed in the LC (Figure 2d), with their orientations isotropic within the plane perpendicular to $\mathbf{n_i}$ (Figure 4f). The analyzer **A** placed in the detection channel right before the avalanche photo diode (APD) allows us to select a specific linear polarization of fluorescence light emerging from the sample and then perform the fluorescence decay measurements for different orientations of linear polarization directions of the excitation and emission light (Figure 4f). For example, the slower decay component of the fluorescence signal manifests itself when light detected by the APD is passed through an analyzer $\mathbf{A}\|\mathbf{n_i}$ (Figure 4). When the composite is switched with voltage $U = 10$ V while keeping $\mathbf{A}\|\mathbf{n_i}$ (along the initial rubbing direction), the faster component of fluorescence decay dominates (Figure 4b). As the switching voltage is gradually increased starting from 0 V, the relative contributions from the faster and slower components change gradually, as evident from the continuous variations of amplitudes $A_1$ and $A_2$ used for the double-exponential fit (Figure 4e). At $\mathbf{A}\perp\mathbf{n_i}$, the particles orientation is isotropic with respect to **A**, leading to an increase in the faster decay component arising from the GSDs oriented in the direction of **A**, which is relatively independent of the switching voltage (Figure 4b). Although the fluorescence decay of the GSD-LC composite remains relatively independent of the polarizations of the incident excitation beam, small change in the decay profiles are visible when **P** and **A** are oriented orthogonally to each other, as can be seen from comparing the decay characteristics at $\mathbf{P}\|\mathbf{n_i}, \mathbf{A}\|\mathbf{n_i}$ and at $\mathbf{P}\perp\mathbf{n_i}, \mathbf{A}\|\mathbf{n_i}$ (Figure 4b). The orientation of incident light polarization $\mathbf{P}\perp\mathbf{n_i}$ in this case contributes to a faster fluorescence decay, apparent even when **A** is aligned to collect the slower component of the fluorescence decay in both cases, which is partially due to the relatively low absolute value of the scalar order parameter of GSDs (*i.e.* due to the relatively broad distribution of GSD orientations, Figure 2d).

Our study shows that fluorescence properties of the pre-engineered composite materials, including both the intensity and decay rates of fluorescence, can be controlled by weak external stimuli. These effects can be enhanced through increasing the absolute values of the scalar order parameter of rod-like GSDs for both the orientations perpendicular to the LC director and along. In particular, the inset of Figure 2d demonstrates how such scalar order parameter $S_{GSD}$ can be varied by controlling the strength of surface anchoring and the geometric parameters of GSDs for the case of spontaneous alignment perpendicular to the director **n**. The control of scalar order parameter would allow for narrowing the angular distributions of GSD orientations and, thus, for further enhancing the capability of modulating both the intensity and the decay rates of fluorescence from the LC-GSD composites. While the Purcell effect was demonstrated in many fundamental single-particle-level studies, the possibility of controlling it with low-voltage fields was never demonstrated and its practical uses in realizing composite materials remain scarce.[23-27] Our study shows the possibility of pre-engineering and realizing highly tunable and reconfigurable fluorescence properties of composite materials by combining anisotropic metal-dielectric-dye core-shell colloidal nanostructures[19-22,28] with the facile response and intrinsic orientational ordering of the LC host.

**CONCLUSION**

To summarize, we have designed and practically realized a composite material with tunable intensity and decay rates of fluorescence and comprising ordered gold-silica-dye nanorods dispersed in a nematic LC host. We characterized its unique tunable emission properties that can be controlled by low-voltage fields. The experimental platform we have developed can be extended to optical[29] and other types of weak-stimuli control of the composites. This platform may enable realization of electrically tunable SPASERs, single-photon-emitters and a host of other nanophotonics devices[30,31] that can be enabled by combining facile switching of LCs and other soft matter media with the capabilities of plasmonic nanoparticles to strongly alter localized light-matter interactions. Since the fabrication of devices based on LC-GSD composites is similar to that used in display and electro-optic applications and since our approach can be easily adapted to using not only 5CB but also a variety of other commercially available LC host materials, we expect that the LC-GSD mesostructured composites will enable a large variety of practical technological applications.

**MATERIALS AND METHODS**

**Hybrid GSD Nanostructures Synthesis.** The hybrid GSD nanostructures were prepared using a modified method of coating silica layer on hexadecyltrimethylammonium bromide (CTAB, Sigma-Aldrich)-stabilized GNRs.[28] The as-prepared GNRs obtained using the seed-mediated method[32] were centrifuged twice at 9000 rpm for 20 min (Sorvall RC5C centrifuge, Sorvall Heraeus) to remove excessive CTAB and other redundant chemicals. Then 10 µL of 0.2 M CTAB was added into 2 mL of aqueous dispersion of GNRs with optical density (OD) of 3 to reach a proper concentration of CTAB, which acts as structure-directing agent for the silica shell formation on the surface of GNRs.[33] Then 20 µL of Oxazine 725 perchlorate (Exciton Co., 11.5 mg/mL, dissolved in deionized water *via* ultra sonication and heating) and 20 µL of a 0.1 M aqueous NaOH solution were added to the GNRs solution under vigorous stirring, followed by three additions of 12 µL of 20 vol.% tetraethyl orthosilicate (Sigma-Aldrich) in methanol under gentle stirring with a 1 h interval, resulting in direct wrapping of dye molecules into the silica layer. After 14 h of reaction, the mixture was centrifuged twice at 6000 rpm for 10 min (Sorvall Legend Micro 21 Centrifuge, Thermo Scientific), and the resultant

GSDs were re-dispersed in deionized water. The distributions of GSD and GNR dimensions are characterized in the Figure S1.

**Dispersion of GSD Nanostructures in a Nematic LC.** To achieve homeotropic alignment of LC molecules at the surface of GSDs, these nanostructured particles were further capped with DMOAP (60% in methanol, ACROS Organics): 4 mL of GSDs solution with OD of 2.3 was diluted to 8 mL and then 400 µL of DMOAP was injected under stirring. The solution was kept stirred for 20 min and washed by four cycles of centrifugation at 6000 rpm for 10 min and transferred into methanol.[15] The DMOAP-coated GSDs solution that resulted from this procedure was then mixed with 15 µL of a nematic 5CB, (TCI Co., Ltd.) and the solvent was fully evaporated at 65 °C for 1 h. The dispersion in isotropic phase of 5CB was then sonicated at 40 °C for 5 min, followed by a vigorous stirring while the sample was being cooled down to nematic phase and followed by centrifugation at 2000 rpm for 5 min, yielding a stable colloidal dispersion of individual GSDs in the LC host.[16]

**Planar LC Cells Fabrication.** Planar LC cells were prepared from glass slides with inner surfaces coated with ITO transparent electrodes. These ITO-glasses were further spin-coated by a thin layer of polyvinyl alcohol (PVA, 1 wt.%) at 3000 rpm for 30 s, baked at 100 °C for 1 h, rubbed along a single direction for three times using velvet cloth in a home-made rubbing setup to impose the unidirectional planar boundary conditions for the LC director **n**, defining initial uniform ground-state orientation **n$_i$** of the director in a monocrystal sample at no applied fields. The substrates were then glued together with a UV-curable glue NOA-65 (Norland Products, Inc.) containing 30 µm silica spacers to define the thickness of the LC cell. These surface boundary conditions are strong, so that the director orientation at the confining surfaces is fixed along **n$_i$** even at rather large applied fields, when the director orientation **n(r)** is changing within the sample bulk. The cells were infiltrated with the LC-nanoparticle composite using capillary action and subsequently sealed at cell edges using fast setting epoxy.

**TEM, Optical Microscopy and Fluorescence Lifetime Measurement Setup.** TEM imaging was performed using a FEI Philips CM 100 microscope after drop casting the dilute solutions of GSD on a copper grid. Optical polarized extinction spectra were obtained using Olympus BX51 microscope combined with a spectrometer (USB 2000, Ocean Optics Inc.). The electro-optic response of the GSD-LC composites was characterized using a Si amplified photodetector PDA100A (Thorlabs Inc.) and a multifunction data acquisition system USB-6363 (National Instruments Co.) controlled by a homemade program written in Labview (National Instruments Co.). Characterization of fluorescence from GSDs dispersed in the LC was performed using an inverted optical microscope Olympus IX81. We used green light from a supercontinuum source to excite the nanoparticles, which matches the spectral location of the transverse surface plasmon resonance mode of the GSDs. To obtain this excitation light, the 780 nm output from a Ti:Sapphire oscillator (140 fs, 80MHz repetition rate, Chameleon Ultra II, Coherent) pumped a highly-nonlinear, polarization-maintaining photonic crystal fiber (FemtoWhite-800, NKT Photonics) to generate the supercontinuum output, which was then sent through a band pass filter (510-560 nm) and a $\lambda/2$ wave plate to control the polarization. For fluorescence decay experiments, the samples were illuminated through a 100× objective (Numerical aperture 1.4, from Olympus) and fluorescence signal from the samples were detected using the same objective in the epi-detection mode. The optical signals were sent through suitable filters and analyzer before being detected by an APD (PicoQuant, τ-SPAD) connected to a time correlated single photon counting device (TCSPC, Becker and Heckle, SPC 130). The repetition rate of the excitation source (80 MHz) allows us to measure the fluorescence decay within the time range of 12.5 ns, sufficient to study the fluorescence dynamics of dye molecules. The polarized fluorescence spectra were measured by exciting a typical sample area ~100 $\mu m^2$ with light from the same excitation source using a 20× objective and then collecting the emitted light through a 40× objective in the forward detection geometry and sending it to the spectrometer through an analyzer.

**Simulation of Electric Field in GSD Nanostructures Based on Discrete Dipole Approximation (DDA) Method.** An electric field distribution around GSD nanostructures was calculated using a DDA method.[34-37] We utilized an online computational tool available at nanohub.org[38] for calculating the electric field pattern around GNRs in the central parts of GSDs. Three-dimensional (3D) structures of silica capped GNRs were drawn using a 3D modeling software (Blender) and rendered into the DDA simulation software (DDSCAT) to form a corresponding set of dipoles, with the spatial distribution of refractive index based on the material properties of gold and silica. The calculations were performed with 1 dipole/nm$^3$ resolution. In order to match the experimental conditions, we simulated the extinction spectra of GSD and compared them with experimental spectra of GSDs in water. Fine adjustments were made in the size of gold nanostructure and thickness of silica shell to match the experimental SPR peak position as necessary. We then calculated the electric field pattern ($|E|^2$) around GSD for polarizations of the incident beam parallel and perpendicular to the long axis and wavelength corresponding to excitation and emission lines (Figure 1c-f), matching the experimental conditions.

**ASSOCIATED CONTENT**

Supporting Information

## AUTHOR INFORMATION

**Corresponding author**

*Email: ivan.smalyukh@colorado.edu.

**Author Contributions**

L.J. and Q.L. synthesized nanoparticles. H.M. did numerical modeling of the electric field enhancement due to GNRs within the GSD particles. L.J., H.M., Q.L. and I.I.S. performed experimental work and analyzed results. All authors contributed to writing the manuscript. I.I.S. conceived and designed the project.

**Notes**

The authors declare no competing financial interest.

## ACKNOWLEDGMENTS

The authors thank G. Al Abbas, P. Ackerman, Y. Yuan, T. Lee and B. Senyuk for discussions. This research was supported by the U.S. Department of Energy, Office of Basic Energy Sciences, Division of Materials Sciences and Engineering, under Award ER46921, contract DE-SC0010305 with the University of Colorado Boulder. The electric field enhancement due to GNRs within GSDs was performed by using an online tool based on DDA method (nanoDDSCAT) available at nanohub.org.

# Figure Captions

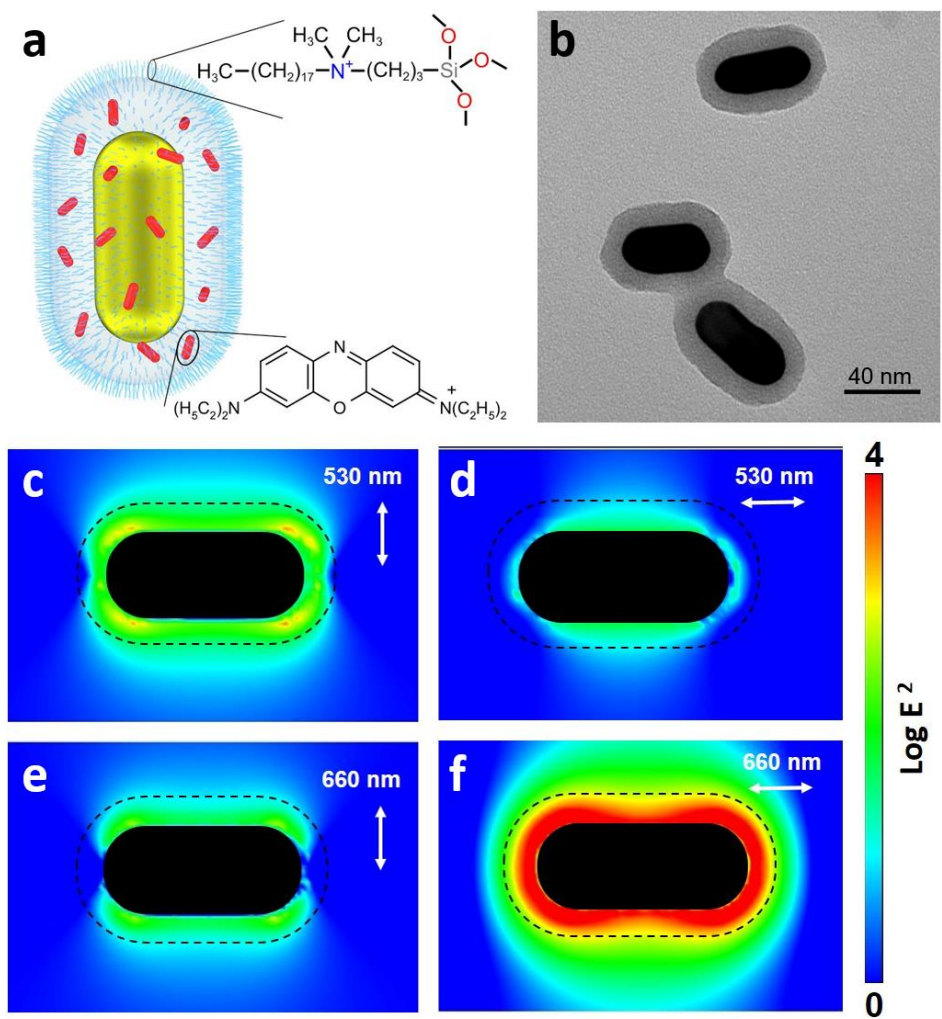

**Figure 1.** GSD nanostructure and its SPR-enhanced electric field. (a) Schematic of the GSD nanoparticle with the fluorescence dye Oxazine 725 (red cylinder) embedded in a silica shell that is further covered with homeotropic surface alignment layer of DMOAP. (b) TEM image of GSDs. (c-f) Computer-simulated electric field enhancement within a GSD nanostructure for different polarizations of excitation light in the vicinity of transverse (530 nm) and longitudinal (660 nm) SPR peaks of the GNR core, respectively. The color-coded scale in the bottom-right of figure shows the relative values of $|\mathbf{E}|^2$.

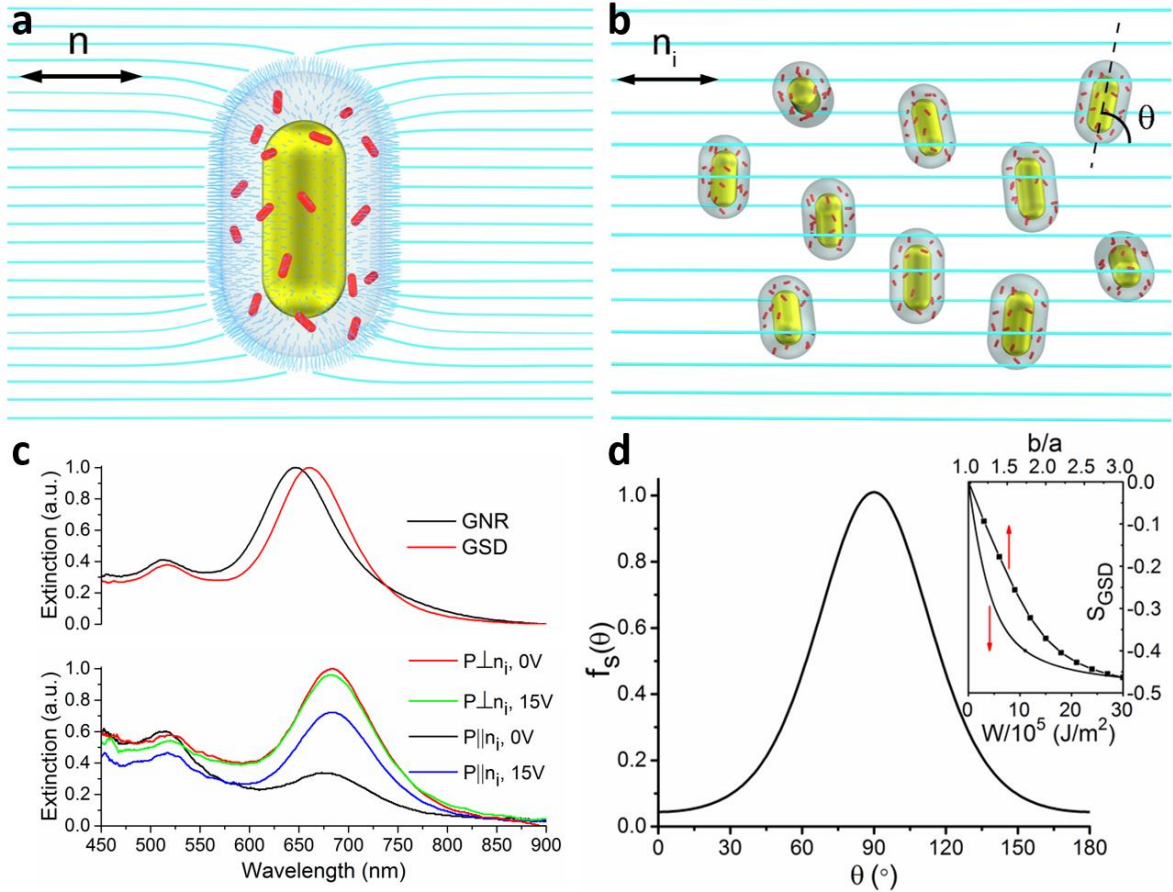

**Figure 2.** Alignment and electric switching of GSDs in a nematic LC. (a) Schematic of a GSD with weak homeotropic boundary conditions in a nematic LC. (b) Schematic of dispersed GSDs in nematic LC with the uniform far-field alignment $\mathbf{n_i}$, with the definition of angle θ between the long axes of GSDs and $\mathbf{n_i}$. (c) Top: Extinction spectra of GNRs and GSDs; Bottom: Voltage-dependent extinction of GSDs in a nematic LC for **P** orientations along and perpendicular to the rubbing direction. The voltage dependent SPR spectra of GSDs are consistent with those of GNRs with similar LC-induced alignment that we studied in Ref. 15. (d) Computer-simulated distribution of GSD orientations with respect to $\mathbf{n_i}$ at $S_{GSD}$ = -0.28. The inset the control of $S_{GSD}$ by varying $W$ and the ellipsoidal GSD aspect ratio $b/a$.

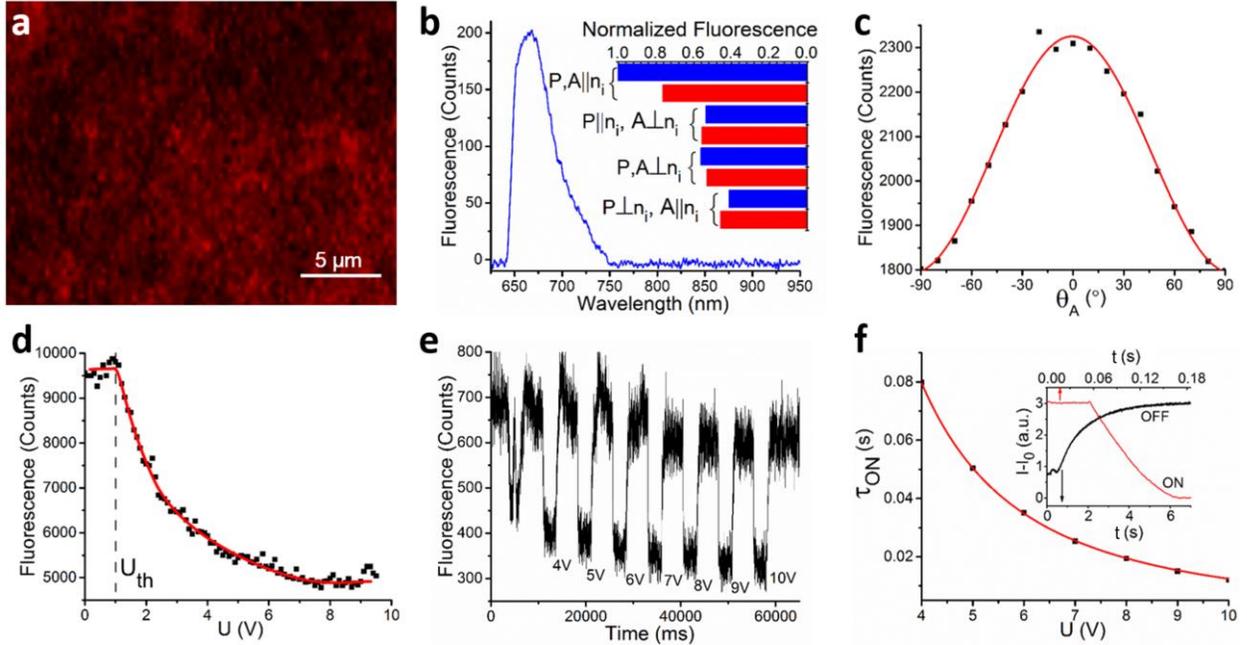

**Figure 3.** Electric switching of SPR-enhanced fluorescence intensity of GSDs in a nematic LC. (a) A fluorescence micrograph showing the GSD-LC composite in a planar glass cell. (b) Fluorescence spectrum of GSDs in nematic LC. The inset shows normalized fluorescence intensity detected for different combination of **P** and **A** and applied voltages $U$ (blue and red bars correspond to $U = 0$ V and $U = 10$ V, respectively). The fluorescence intensity is maximized when the absorption efficiency is at maximum, which is for the polarization of excitation light perpendicular to rods (transverse SPR) and along $\mathbf{n_i}$ and for the analyzer parallel to the excitation polarizer, so that the maximum fraction of the emitted polarized fluorescence can be detected. (c) Polarization-dependent fluorescence intensity for $\mathbf{P}\|\mathbf{n_i}$. (d) Voltage-dependent fluorescence of GSDs in a nematic LC. (e) Switching of GSDs' fluorescence intensity by an electric field. (f) Voltage-dependent ON and OFF switching times obtained from the change of transmission $I$-$I_0$ versus time curves, such as the one shown in the inset for $d = 30$ μm and $U = 4$ V. The red curve is a fit of the ON time-voltage dependence,[18] $\tau_{ON} = \tau_{OFF}/[(U/U_{th})^2 - 1] = 2.78/[(U/0.668)^2 - 1]$.

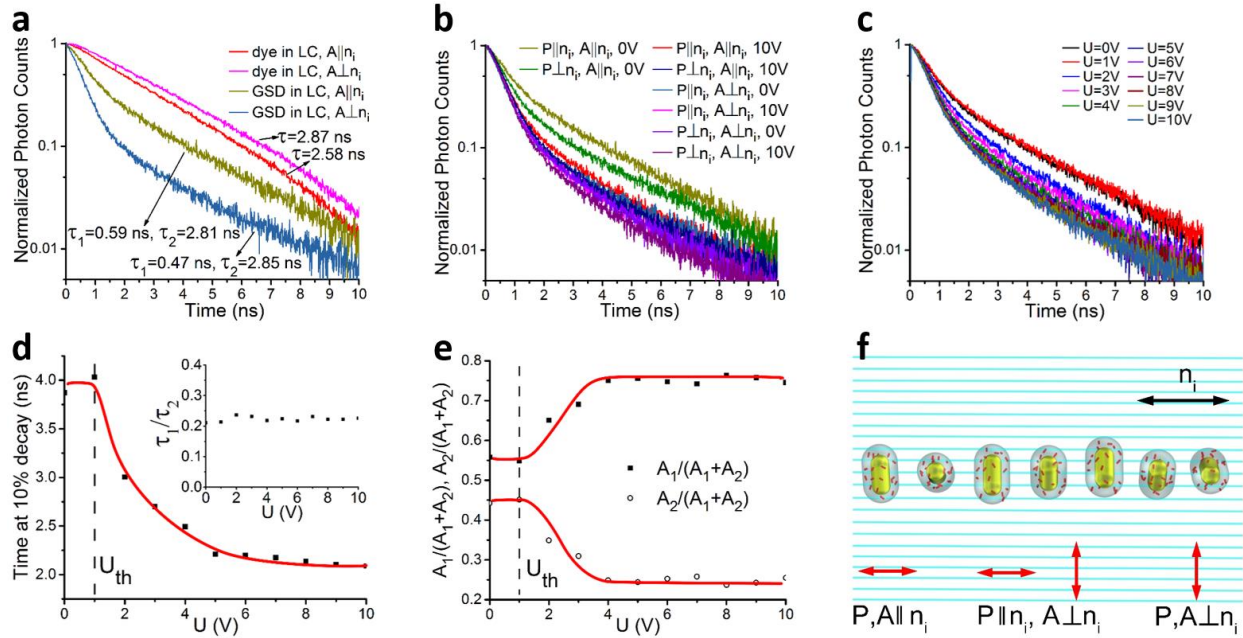

**Figure 4.** Electric switching of fluorescence lifetime of GSDs in LC. (a) Polarization-dependent fluorescence intensity decay of Oxazine 725 and GSDs in a nematic LC for $\mathbf{P}\|\mathbf{n_i}$; (b) Polarization- and voltage- dependent fluorescence decay of GSDs in a nematic LC. (c) Voltage-dependent fluorescence intensity decay of GSDs in a nematic LC for $\mathbf{P}\|\mathbf{n_i}$ and $\mathbf{A}\|\mathbf{n_i}$. (d) Voltage-dependent decay time of GSDs corresponding to the fluorescence count decrease to 10% of its original level. The inset shows that the ratio $\tau_1/\tau_2$ is roughly voltage-independent. (e) Voltage-dependencies of $A_1/(A_1+A_2)$ and $A_2/(A_1+A_2)$. (f) Schematic of GSDs in nematic LC showing the origin of the dependence of fluorescence lifetime on the polarization of light selected by $\mathbf{P}$ and $\mathbf{A}$ in the excitation and detection channels, respectively.